# Temperature Dependence of the Anisotropy Field of $L1_0$ FePt near the Curie Temperature

*H.J. Richter and G.J. Parker*

*Western Digital Corporation*

*1710 Automation Parkway, San Jose, CA, 95131*

*Abstract*

Near the Curie temperature the anisotropy field of magnetically uniaxial $L1_0$ FePt is expected to follow the scaling law $(1-T/T_c)^\beta$ where $T$ is the temperature and $T_c$ the Curie temperature. In the literature β values between 0.36 and 0.65 have been reported. Based on recording measurements and micromagnetic analysis, we show that only the values of β near the low end of the reported range are compatible with the data. We also conclude that thermally activated magnetization reversal at temperatures near $T_c$ cannot be ignored, even at time scales smaller than 1 ns. We demonstrate that thermally activated magnetization reversal at temperatures close to $T_c$ is well described by conventional theory with a frequency factor $f_0$ of the order of $10^{12}$ Hz. It is reasoned that the unusually high value for $f_0$ is a consequence of the temperature-induced reduction of the degree of alignment of the micro-spins within the grains.

**Introduction**

To further advance the storage density of magnetic recording, media consisting of very small grains are needed. Stable magnetization in these very small grains requires materials with extremely high anisotropies, which can no longer be switched with available magnetic fields thus necessitating a write assist. To date, the most promising write assist scheme is heat-assisted magnetic recording (HAMR), where the recording medium is temporarily heated above the Curie temperature $T_c$. The information is written during the cooling process at temperatures $T_{wr} < T_c$ where the anisotropy is reduced and available write fields can switch the magnetization.

For HAMR, L1$_0$ FePt is the material of choice due to its relatively low $T_c$ and very high anisotropy. To understand the writing process, it is crucial to have exact knowledge of the temperature dependence of the saturation magnetization $M_s$ and the anisotropy field $H_A$. Since the Curie point represents a second order phase transition, one expects equations of the following form for the magnetization $m(T)$ and the anisotropy field $h_A(T)$:

$$m(T) = \frac{M_s(T)}{M_{s0}} \left(1 - \frac{T}{T_c}\right)^\alpha \quad (1),$$

$$h_A(T) = \frac{H_A(T)}{H_{A0}} \left(1 - \frac{T}{T_c}\right)^\beta \quad (2).$$

Here, the index "0" refers to the respective values at zero Kelvin. In the literature, $\alpha$ has been reported to be 0.324 [1] and 0.35 [2], which is regarded as good agreement within numerical accuracy. On the other hand, the anisotropy constant $K(T) = 0.5\mu_0 M_s(T) H_A(T)$ was reported to approximately follow $(1-T/T_c)$ which implies that $\beta \approx 0.65$ [2]. This is contrasted with the result in [1], where $K(T)$ approximately follows $(1-T/T_c)^{0.66}$ or $\beta \approx 0.34$. Myrasov et al [3] have shown that $K(T) \propto M_s(T)^{2.1}$, where the unusual exponent of 2.1 is caused by the domination of two-ion anisotropy in L1$_0$ FePt. This result is incompatible with ref [2], but in agreement with [1]. Although not stated explicitly, it appears that the data shown in [3] for $K(T)$ imply $\alpha \approx 0.43$ and $\beta \approx 0.47$, which disagrees with the $\alpha$ values reported in refs [1] and [2].

Anisotropy and magnetization measurements on thin films of L1$_0$ FePt have been reported in [4], where the relation $K(T) \propto M_s(T)^2$ was originally found. The temperature dependence of $K$ is approximately $(1-T/T_c)^{0.72}$ and is roughly compatible with [1]. Since it is well known that $T_c$ and the anisotropy of L1$_0$ FePt depend on grain size [5], it is highly desirable to have experimental data on granular systems rather than thin films. In this case, standard magnetometry yields no useful information because the grains are superparamagnetic in the temperature range of interest at experimentally accessible time scales. This leaves recording measurements as the best approach. One such study has been conducted [6] and it was found that the critical exponent β is about 0.52, which implies α ≈ 0.47 if $K(T) \propto M_s(T)^{2.1}$ is valid.

Thus, there is a clear need to identify which of the reported dependencies apply. From an application point of view, the detailed knowledge of $h_A(T)$ is important, because the slope $dh_A(T)/dT$ relates to the effective write field gradient [7] and a stronger temperature dependency (i.e. smaller β) is advantageous. In the following, we will tackle the problem by recording measurements of a different nature as reported in [6] and will interpret the results by a micromagnetic simulation of the experiment. For the interpretation of the data, three different β values will be considered: 0.3564 (which is an improved estimate of the work in[1] and in the following referred to as 0.36), 0.5 and 0.65.

## Experimental

Fig.1 shows a sketch of the temperature of the medium when it passes by the recording head. As indicated, the maximum temperature surpasses $T_c$. According to equation 2, the anisotropy field varies as shown, where we assign zero if $T > T_c$. The recording takes place at the write temperature $T_{wr} < T_c$ at the trailing edge of the thermal write bubble on the right hand side in Fig. 1. $T_{wr}$ depends on the applied field $H_{wr}$ and is indicated by the short horizontal line at the trailing edge. In this work, we assume that the write field is constant near the write location, where the justification for this assumption will become evident later. Since the direction of the write field $H_{wr}$ is at an angle $\vartheta_0$ to the easy axis of the recording medium we apply the concept of an effective field $H_{wr,eff} = H_{wr}/h_{SW}(\vartheta_0) = H_{wr} \cdot (\cos^{2/3}\vartheta_0 + \sin^{2/3}\vartheta_0)^{3/2}$ as it follows from the Stoner-Wohlfarth model [8].

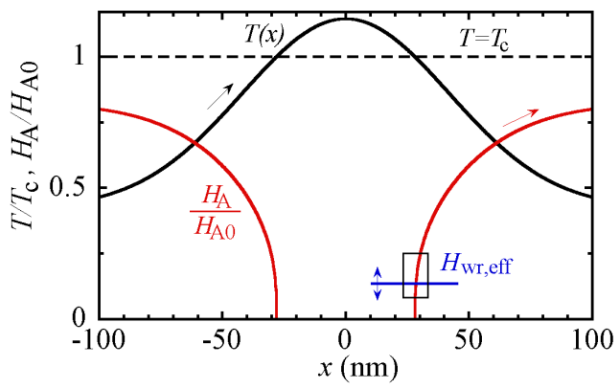

Fig 1. Illustration of the writing process for HAMR. With thermal activation neglected, writing occurs during cooling at the write temperature $T_{wr} < T_c$ where the anisotropy field $H_A$ surpasses again the write field $H_{wr}$ (head field). Changing the applied field traces out a part of the $H_A(T)$ curve as exemplified by the box.

As can be seen in Fig. 1, increasing the write field magnitude moves the write point further downstream to lower temperatures. Therefore, by changing the write field, a portion of the curve $h_A(T)$ can be traced out as indicated by the box in Fig. 1. This idea was put forward in [6], where the change of the write field was accomplished by modulating the write current magnitude resulting in a measurable shift of the written transitions. Here, we made a conscious decision not to apply this scheme because the recording field does not change instantaneously when a transition is being written and it becomes problematic to assign a single well-defined write field magnitude to this process.

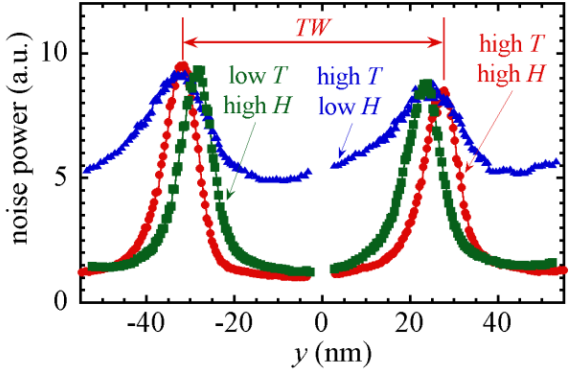

*Fig. 2: Examples for the track width measurement. The track width TW is the distance between the peak locations of the noise powers as explained in the text. Both increasing the temperature and increasing the write field increases the track width, where the temperature effect is much more pronounced. For very low fields, writing is incomplete as shown by the curve "high T low H".*

Instead, we map out the write locations in the cross-track direction by measuring the (physical) track-width at a very low linear density. Then the read-back signal is dominated by the center portion of the recorded magnets and unaffected by field rise time effects. Measuring the track-width cannot be accomplished by standard track-scans because there are various read-back phenomena (track curvature, flux closures at the track edges) that lead to deviations between the actual track-width and the reported full-width half maximum [9]. A better way to measure track width is to make use of the fact that zero net magnetization has the highest magnetization noise [10]. To use this effect, two low-density recordings are made at a distance of about +/-65% of the anticipated track width and subsequently another low-density recording is made in the center. The center track partially erases the outer tracks and the magnetization of the center track is opposed to that of the outer tracks for about 50% of the track length. Consequently, the magnetization makes a transition in the cross-track direction and creates a noise strip at the track edge. This noise strip can be detected and results in a peak of the noise power at either track edge. The procedure was simulated using micromagnetic modeling and it was verified that the distance between the noise peaks is identical to the width of the written magnetization, which is defined as the distance where 50% of the grains can be switched. The noise peaks can be considered isolated as long as the written track is at least 1.5 times wider as the reader width.

To map out $h_A(T)$ the track width is measured for a variety of write currents and laser powers. It is noted that a change in either laser power or write current causes a change in writer protrusion and an adjustment of the writer heater power is necessary to keep the clearance between the head and the disk constant during writing. Fig. 2 shows three examples of these noise measurements: one at high laser power and high write current, one at low laser power and high write current, and one at high laser power and low write current. As can be seen, the tracks become wider with increasing laser power and/or write current, where the effect of the laser power is considerably stronger than that of the write current. (This justifies the previous assumption to consider constant write field for the determination of the writing location). It can also be seen that the noise power for the lowest write current is significantly higher at the track center, which means that the applied field is not strong enough to switch all grains. This is in accord with the micromagnetic simulations of the experiment.

## Results

For the remainder of the paper, the data are parameterized by the ratio $\eta = H_{wr,eff}/H_{A0}$, that is, the ratio of the effective write field to the anisotropy field at zero Kelvin. For the media used here, we estimate $\mu_0 H_{A0}$ to be 9T. Head modeling shows that the write field is proportional to the write current to first order. At high currents, slight saturation effects occur which we take into account.

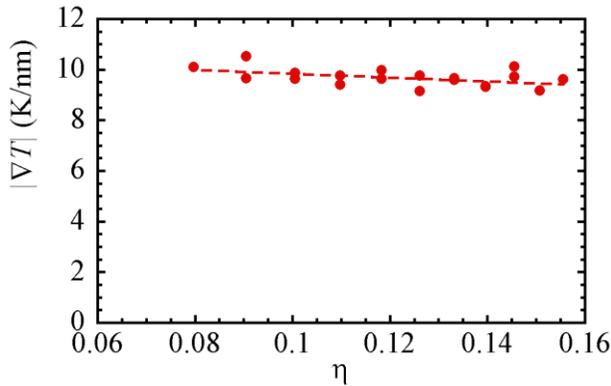

Fig. 3: Measured cross-track gradient as a function of the field ratio $\eta = H_{wr,eff}/H_{A0}$

A typical set of experimental data consists of 19 different write fields and 13 different laser powers for each write field. For constant write field, the change in track width caused by the laser power change can be used to find the thermal gradient in the cross-track direction $y$ [11]:

$$\frac{dT}{dy} = \frac{\delta P}{P}\frac{\Delta T_{wr}}{\Delta y} \qquad (3).$$

Here $\delta P/P$ is the relative laser power change and $\Delta T_{wr} = T_{wr}-T_a$ where $T_a$ is the ambient temperature. Equation (3) is applied to each write field. Fig. 3 shows an example and it can be seen that the thermal gradient increases somewhat with lower write field, which is precisely what is indicated in Fig. 1. Equation (3) can be integrated to give the total temperature change $\delta T_{wr}(\eta)$ that occurs when the write field is changed:

$$\delta T_{wr}(\eta) = \int \frac{\partial T}{\partial y} dy + C \qquad (4).$$

Equation (4) has an unknown integration constant $C$ which means that we do not know where to place the box in Fig. 1.

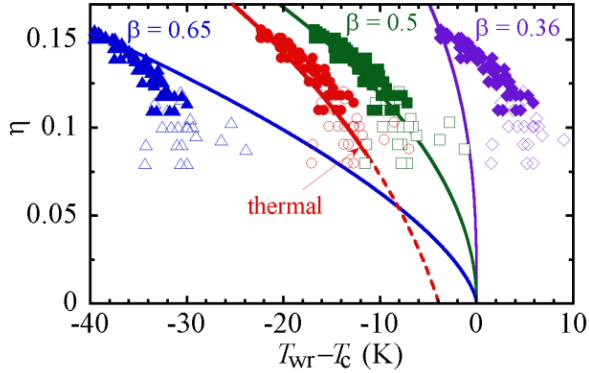

*Fig. 4: Comparison between the experimental data and the scaling law (equation (2)) for the anisotropy field as a function of temperature using the various β values. The open symbols indicate that the dcSNR is less than 15dB. The curve labeled "thermal" is equation (5) with grain size 6.5nm medium thickness 10nm, $\mu_0 H_{A0} = 9T$, $M_{s0} = 1150$ kA/m and damping 0.05. The "thermal" curve is shown dashed when the dcSNR is predicted to be smaller than 15dB.*

If thermally activated magnetization reversal is ignored, the write temperature is defined by equation (2). For convenience, for each β, equation (2) is plotted as function of $T_{wr}-T_c$ using the notation $\eta = H_{wr,eff}/H_{A0}$. The measured data can directly be compared to the theory by demanding that one data point (we have chosen the highest η value) must fall on its respective theory curve, thus defining the integration constant $C$. Accordingly, as shown in Fig. 4, the same data appear three times for the three β-values under discussion. It can readily be seen that either β = 0.36 or β = 0.65 fit the data and we therefore come to the conclusion that β = 0.5 fits the data best among the considered choices. This is the same result that was obtained in [6].

It was mentioned before that the media can not be saturated at low write fields. Then only a fraction of the grains participate in the switching process resulting in an average switching field lower than that of the entire ensemble and these data points should be given less consideration. To quantify this effect, the signal-to-noise ratio (SNR) for the dc-magnetized magnets is also measured and all symbols are shown filled if the on-track dcSNR is greater than 15dB.

## Thermal Activation

It is important to check whether it is legitimate to ignore the effect of thermal activation. The experiment is simulated with micromagnetic modeling where the thermally activated magnetization reversal of the grains at high temperatures is described by the stochastic Landau-Lifshitz-Bloch equation derived in [12]. The modeling involves multiple head designs and each data point represents an individual grain. The properties of the grains are distributed and result in the data clouds shown. In Fig. 5, the various scenarios for the three β values 0.36, 0.5 and 0.65 are summarized and plotted in the same fashion as in Fig. 4. For each β, the lines indicate the limit given by equation 2. It has been verified that the calculations successfully retrieve equation 2 when both the stochastic and the demagnetizing effects ("NSD") are removed. The "spines" that are visible on some of the data clouds are due to the finite spatial resolution used and can be disregarded. The filled symbols represent the results of the calculations with all effects included and are the ones to be compared with experiment. Evidently, the final result for β = 0.36 is similar to the limiting case for β = 0.5, especially for temperatures not too close to $T_c$. The open symbols show the results where the stochastic fields ("NS") have been removed. Both the demagnetizing and the stochastic fields make the curves η($T_{wr}$–$T_c$) appear shallower, but the stochastics clearly dominate the effect. It is therefore concluded that thermally activated magnetization reversal cannot be ignored with the implication that β = 0.36 has to be used in equation (2). In other words, the resemblance of the data to equation (2) for β = 0.5 has no direct physical meaning.

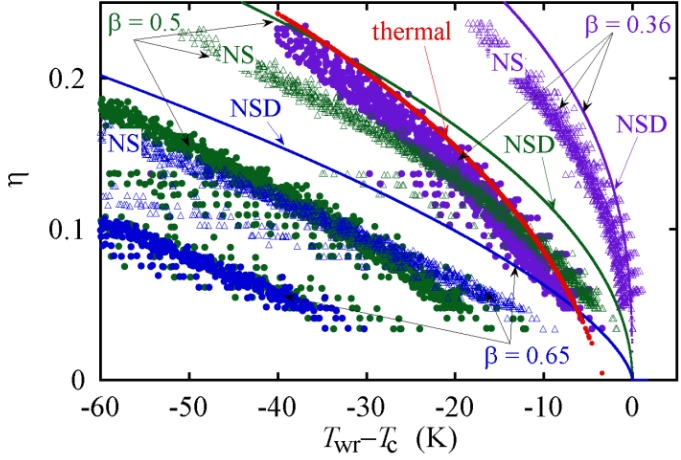

*Fig. 5: Micromagnetic results for the various scenarios β = 0.36, 0.5 and 0.65. For each β value, three curves are given, where the full line is the scaling according to equation 2 ("NSD"). The open symbols correspond to simulations with the stochastic fields removed ("NS") and the demagnetizing fields included. The filled symbols correspond to the full simulation with all effects taken into account. For comparison, the curve labeled "thermal" shows the result obtained with equation (5). Parameters: $\mu_0 H_{A0}$ = 7.2T, D = 7.7nm and $M_{s0}$ = 1200kA/m; all other parameters are the same as in Fig. 4.*

In the last section of the paper we show that thermally activated magnetization reversal can be successfully described by the conventional Arrhenius-Neel formalism even if the temperature is near $T_c$. It is well known that thermally activated magnetization reversal is described by a relaxation time $\tau = 1/f_0 \exp(-\Delta E(h)/kT)$, where $\Delta E$ is the (field dependent) energy barrier between the two stable magnetization states, $h$ is the applied field normalized to the anisotropy field $H_A$, and $k = 1.38 \times 10^{23}$J/K is Boltzmann's constant [13]. This can be used to find the time dependent switching field of single domain particles that reverse magnetization according to the Stoner-Wohlfarth model [14]:

$$H_{sw}(\vartheta_0) \cong H_A h_{SW}(\vartheta_0)\left[1 - \left(\frac{kT}{KV}\frac{\ln f_0 t}{\ln 2}\right)^{2/3}\right] \qquad (5).$$

Here $V$ is the particle volume and $h_{SW}(\vartheta_0)$ is the angle dependence of the switching field according to the Stoner-Wohlfarth model as already given further above. Equation (5) is a good approximation if $\vartheta_0$ is not close to 0 or $\pi/2$. If the applied field is along the easy axis, $\vartheta_0 = 0$, the classical frequency factor $f_{0,class}$ is [13]:

$$f_{0,\text{class}} = \frac{\alpha_D}{1+\alpha_D^2} \mu_0 H_A \gamma \sqrt{\frac{KV}{\pi kT}} (1-h)^2 (1+h) \tag{6},$$

where $\alpha_D$ is the damping constant and $\gamma=1.76\times10^{11}$ T/s the gyromagnetic ratio. Similar equations have been shown to fit numerical data if $\vartheta_0$ is not close to zero [15].

It is well known that equation (5) breaks down for short times, when $f_{0,\text{class}}t$ approaches 1 and the field required to switch the magnetization increases sharply [14]. For our application, with typical grain sizes $D$ around 7 nm and a linear velocity of $v = 20$m/s, the field (and temperature) exposure time is $t_{\exp} = 2/\pi\, D/v$ and of the order of 0.2ns. This yields $f_{0\text{class}}t_{\exp} < 10$, which means that the validity of equations (5) and (6) is highly questionable.

Importantly, the derivation of equation (6) assumes that the atomic spins within the single domain particle are perfectly aligned and the spin ensemble can be replaced by an equivalent "macro-spin". For high temperatures, the magnetization is highly non-uniform and correspondingly reduced magnetizations and anisotropies are assigned to this macro-spin. In other words, only the aligned fraction of the magnetization is considered and the remaining random part is completely ignored. Therefore, the theory of thermal activation of Brown [13] has to be extended to non-uniform magnetization. To our knowledge, such an extension has not been reported in the literature. It is noted that thermally induced non-uniform magnetization reversal processes have been studied [16], but in these studies the magnetization is only non-uniform at the instant when it crosses the energy barrier, which is a completely different case.

In the following, it is assumed that the basic mathematical form of the relaxation remains valid also for non-uniform magnetization. An inspection of equation (6) shows that $f_0$ becomes small for fields close to the anisotropy field ($h\to1$), which means that the magnetization response becomes sluggish when its instability point is approached. At high temperatures, where the micro-spins are swirled around by the thermal energy and the magnetization is highly non-uniform, it is physically not plausible that such a system would show a sluggish behavior, which suggests that the frequency factor $f_0$ should be modified.

To develop this further, consider one *micro-spin i*. On average, the orientation of this *micro-spin* will make an angle $<\vartheta>$ with the applied field as predicted by the Stoner-Wohlfarth model of the *macro-spin* with correspondingly reduced magnetization and anisotropy as discussed before. In the presence of the thermal energy, the orientation of the *micro-spin i* will fluctuate around $<\vartheta>$ where the fluctuations will increase with temperature. The micro-spin is driven back to its equilibrium by the exchange that it experiences from its neighbors. This exchange field can be estimated by the exchange integral $J$ as $4NN\cdot J/M_{s0}/a^3$, where $NN$ is the number of nearest neighbors and $a \approx c = 0.385$ nm is the lattice constant for L1$_0$ FePt. The exchange integral $J$ is estimated by $3kT_c/[2NN\cdot S(S+1)]$ where $S = 1$. This estimate corresponds to an exchange stiffness at zero Kelvin of $A = 6.3$ pJ/m for $T_c=705$ K. Then the exchange field becomes:

$$\mu_0 H_{ex}(T) = \frac{3kT_c}{M_{s0}a^3} m(T) \tag{7}$$

Here $m(T)$ takes into account that the exchange field is reduced because the surrounding spins are not fully aligned. Following [13], the temporal response of the magnetization scales as $\mu_0 H\gamma$, and we arrive at

$$f_0 = f_{0,class} + \frac{3kT_c\alpha_D\gamma}{a^3 M_{s0}} m(T)(1-m(T)) \tag{8}$$

For $T \ll T_c$, the thermal energy is not strong enough to introduce fluctuations of the orientations of the micro-spins around the equilibrium. This is considered by the additional factor $1-m(T)$ which causes equation (7) to revert back to $f_{0class}$ as it should.

The exchange field $\mu_0 H_{ex}$ at zero Kelvin amounts to roughly 450 T for FePt grains ($T_c = 705$ K, $M_{s0}$=1150 kA/m) and far exceeds the effective field caused by the anisotropy and the Zeeman energies. Owing to the functional form of $m(T)$ the exchange term dominates throughout the entire high temperature range. The frequency factor $f_0$ is then approximately $10^{12}$Hz and almost constant for $0.9T_c < T < 0.99T_c$. For standard magnetic measurements near room temperature, the increase of $f_0$ is small and remains unnoticed.

With these high $f_0$ values $f_0 t_{exp} \approx 250$ and the application of equation (5) is straightforward. It is also noted that the additional field dependence of $f_{0,class}$ complicates the calculation of the dynamic coercivity, but, since the exchange term dominates, the field dependence of $f_{0,class}$ can be ignored. In Fig. 5, the curve labeled "thermal" is obtained using equation (5) for all available combinations of temperature, head field magnitude and angle, with a scaling factor $\beta = 0.36$ for the anisotropy. The micromagnetic data points to be compared are the filled purple symbols labeled $\beta = 0.36$. The agreement with the micromagnetic simulations is excellent and we highlight that *no fitting parameters are involved*. Strictly speaking, equation (5) should be compared to the micromagnetics with stochastic fields included and demagnetizing fields excluded. As outlined before, demagnetization is only a very weak effect here and, above everything else, the agreement would become even better. Using equation (5), an additional curve labeled "thermal" has also been added to Fig. 4. This demonstrates once again that the experimental data are compatible with thermally activated magnetization reversal and an anisotropy scaling factor $\beta = 0.36$.

In summary, we have shown that the temperature dependence of the anisotropy field in L1$_0$ FePt follows equation 2 with the exponent $\beta = 0.36$. We have also shown that thermal activation cannot be neglected at short times and high temperatures. Conventional thermal activation theory can successfully applied but the frequency factor $f_0$ has to be increased to account for non-uniform magnetization in the grains.

## References

[1] A. Lyberatos, D. Weller, and G. J. Parker, *J. Appl. Phys*, **114**, 233904 (2013).

[2] R. H. Victora, and P.W. Huang, *IEEE Trans. Magn,* **49**, 751, (2013).

[3] O.N. Mryasov, U. Nowak, K.Y. Guslienko and R. W. Chantrell, *Europhys. Lett*. **69**, 805 (2005).

[4] J.U. Thiele, K.R. Coffey, M.F. Toney, J.A. Hedstrom, and A.J. Kellock, *J. Appl. Phys*, **91**, 6595 (2002).

[5] C. B. Rong, Y. Li, and J.P. Liu, *J. Appl. Phys*, **101**, 09K505 (2007).

[6] J. Hohlfeld, X. Zheng, and M. Benakli, *J. Appl. Phys*, **118**, 064501 (2015).

[7] T. Rausch, J.A. Bain, D.D. Stancil, T.E. Schlesinger, *IEEE Trans. Magn,* **40**, 137, (2004).

[8] E. C. Stoner and E. P. Wohlfarth, *Philos. Trans. R. Soc. London*, Ser. A 240, 599 (1948).

[9] R.W. Wood and D.T. Wilton, *IEEE Trans. Magn,* **44**, 1874, (2008).

[10] J. Chen and H.J. Richter, US patent US 6166536 A, (2000).

[11] H.J. Richter, C.C. Poon, G. Parker, M. Staffaroni, O. Mosendz, R. Zakai, and B. C. Stipe, *IEEE Trans. Magn,* **49**, 5378, (2013).

[12] M. Tzoufras and M.K. Grobis, *New J. Phys*, **17**, 103014 (2015).

[13] W.F. Brown, *Phys Rev.*, **130**, 1677 (1963).

[14] H.N. Bertram and H.J. Richter, *J. Appl. Phys.*, **85**, 4991 (1999).

[15] X. Wang, H. N. Bertram, and V.L. Safonov, *J. Appl. Phys.*, **92**, 4560, (2002).

[16] H.B. Braun, *Phys. Rev Lett,* **71**, 3557, (1993).